# Analyzing the disciplinary focus of universities:

## Can rankings be a one-size-fits-all?

Nicolas Robinson-Garcia
*EC3metrics spin-off, Spain*

Evaristo Jimenez-Contreras
*Universidad de Granada, Spain*

## ABSTRACT

*The phenomenon of rankings is intimately related with the government interest in fiscalizing the research outputs of universities. New forms of managerialism have been introduced into the higher education system, leading to an increasing interest from funding bodies in developing external evaluation tools to allocate funds. Rankings rely heavily on bibliometric indicators. But bibliometricians have been very critical with their use. Among other, they have pointed out the over-simplistic view rankings represent when analyzing the research output of universities, as they consider them as homogeneous ignoring disciplinary differences. Although many university rankings now include league tables by fields, reducing the complex framework of universities' research activity to a single dimension leads to poor judgment and decision making. This is partly because of the influence disciplinary specialization has on research evaluation. This chapter analyzes from a methodological perspective how rankings suppress disciplinary differences which are key factors to interpret correctly these rankings.*

## INTRODUCTION

In the last five years we have observed a rapid transformation on the way research policymakers use university rankings. These tools have rapidly been integrated as a new support tool on which to base their decisions. They have reshaped the higher education landscape at a global level and become common elements of politicians and university managers' discourse (Hazelkorn, 2011). Not only have they become external key factors as a means to attract talent and funds, but they are also used as support tools along with bibliometric techniques and other methodologies based on publication and citation data (Narin, 1976). Their heavy reliance on bibliographic data has stirred the research community as a whole, raising serious concerns on the suitability of such data as a means to measure the 'overall quality' of universities (Marginson & Wende, 2007). At the same time, university rankings have caught bibliometricians off guard. Although they use them quite often (i.e., journal rankings), they have traditionally disregarded them for institutional evaluation, focusing on more sophisticated techniques and indicators (Moed et al., 1985). On the other hand, university rankings have been traditionally based on survey data and have not considered the use of bibliometric indicators until recently. Moreover, despite their success in the United States, they have had little presence in the European research policy scenario (Nedeva, Barker & Osman, 2014).

The launch of the Shanghai Ranking in 2003 did not only set up the starting point of the globalization of the higher education landscape, but introduced bibliometric-based measures to rank universities. Surprisingly, the Shanghai or the Times Higher Education World University Rankings and QS Top Universities Rankings were not produced by bibliometricians, not even by practitioners. This caught from the beginning the interest of the bibliometric community who rapidly positioned themselves against the use of these tools. Such strong opposition is resumed in the correspondence maintained between Professor van Raan from Leiden University and the creators of the Shanghai Ranking (Liu, Cheng & Liu, 2005; van Raan, 2005ab). Here, van Raan (2005a) highlights serious methodological and technical concerns which are later emphasized by others (i.e., Billaut, Bouyssou & Vincke, 2009). Such shortcomings have to do with the careless use these rankings make of bibliometric data, neglecting many of the limitations bibliometric databases have, and offering compound indicators of dubious meaning which intend to summarize the global position of universities.

Rankings have evolved from marketing tools which have a great impact on the image of universities and their capacity to attract talent and funds (Bastedo & Bowman, 2010) to research evaluation tools which are used strategically by research policymakers shaping their political agenda (Pusser & Marginson, 2013). However, their strong focus on research and their reliance on bibliometric data, entail important threats and misinterpretation issues which may 1) endanger the institutional diversity of universities, and 2) misinform policymakers on the performance of universities or national higher education systems. Considering that most university rankings analyze basically the research performance of universities defined as their publication output (Marginson & Wende, 2007), this chapter discusses the threats the use of university rankings impose to the disciplinary profile of universities. The main thesis is that rankings still offer a restricted view of the research performance of universities, despite the professionalization and rigorousness they have developed in the last few years; converging into research evaluation systems and offering different league tables and a wide range of indicators. Also, the recent trend towards the provision of a wide range of sophisticated indicators (i.e., Centre for Science and Technology Studies, 2015) introduces tensions with the demands of policymakers towards easy-to-use evaluation tools. As a proposal, the use of science mapping and visualization techniques are proposed making use of the information provided by rankings as a means to surpass such tension between what bibliometricians can develop and the demands of research managers.

The aim of this essay is twofold. First, it intends to provide a deep and critical understanding of the methodological decisions implied in university rankings with regard to the type of data used to analyze research and the disciplinary biases the data sources employed have. Such limitation can distort the picture offered by rankings and, if not considered, may lead decision makers to serious misinterpretations with regard to the performance of universities as well as whole national systems. Each university ranking includes its own methodology and bases most of the final score assigned to universities (if not all) on publication and citation data. What is more, these bibliographic data are retrieved from two specific scientific databases, - Web of Science from Thomson Reuters and Scopus, Elsevier, - which have been reported to have a poor coverage towards certain scientific fields (Moed, 2005). It will only make sense to use university rankings when their methodological choices are alienated with institutional goals. However, this is seldom the case; in fact, in most cases it works the other way around. The criteria set by ranking producers will define the strategy of institutions in a race to improve their position (Bastedo & Bowman, 2010).

But, although there is much literature written denouncing these shortcomings (i.e., Buela-Casal, 2007; Aguillo et al., 2010), rankings will continue been used. Researchers and bibliometricians must accept the fact that these tools are demanded by policymakers and that the most reasonable thing to do is to offer correct guidance when using them. Hence the second goal of this chapter: to suggest the use of complementary tools based on science mapping techniques to better comprehend the information provided by rankings. The aim is to develop tools based on the information provided by rankings as well as by these databases that are easy-to-use and interpret and can assist decision makers when using them. Indeed, science mapping and social network analysis can provide deep insight and understanding of the dynamics of an institution, a whole region or the different actors embedded within them (departments, research groups, etc.) (Noyons, 2005). Different visualization solutions are discussed in this chapter along with examples applied to different case studies. They intend to allow the reader to disaggregate and discern different institutional disciplinary profiles which can affect the positioning of universities as well as to deepen on the characteristics that better explain the patterns these maps show.

This chapter is structured into four sections. The first two sections are directed at the first goal of this chapter: building an understanding on the methodological limitations of international rankings. For this, we start by showing how evaluation schemes have evolved, -from national government-driven evaluation systems to international market-oriented rankings or benchmarking systems,- adopting a logic model highly dependent on bibliometric indicators. The objective is to link these two evaluation schemes and how bibliometric techniques and indicators fit within this landscape as well as the policy-driven nature of university rankings. In this section a historical view will be given of the main milestones occurred since the 1980s until the launch of the first international ranking in 2003; the Shanghai Ranking. The second section furthers the argument by analyzing from a bibliometric perspective, the methodological compromises each the main international rankings make when using bibliometric data. It analyzes two central issues: their reliance on the Web of Science and Scopus databases and on bibliometric classifications.

Finally, we present the main issue of this paper: the threat rankings pose to disciplinary diversity and a proposal for overcoming it. Section 3 discusses the effect rankings have on disciplinary differences between universities. It describes the solutions ranking producers have suggested and the misinterpretation issues they raise even when these are methodologically and technically rigorous. These issues do not only deal with the disciplinary profile of universities but also with their organizational structure as well as with the nature of their research (basic vs. applied research). This section offers the main contribution of this chapter, describing the use of the journal publication profile institutional mapping technique. Finally, the main conclusions of the chapter are wrapped up along with a discussion on the main issues that are being raised with regard to the criticisms bibliometric indicators and methodologies receive and the tensions existing between what research policymakers demand and what can be offered by the bibliometric community.

## FROM NATIONAL RESEARCH EVALUATION SYSTEMS TO INTERNATIONAL UNIVERSITY RANKINGS

Despite following different paths, the phenomenon of rankings is intimately related with government interest in fiscalizing the research outputs of universities. This interest extends back to the beginning of the 20th century and has to do with two different factors. First, there has been a shift in the perception of the main mission of universities. Originally conceived of as elite centers for teaching and training, their role as generators of research and scientific knowledge has increased over the last century. Secondly, but related with the first factor, the way research is conducted has also changed. Research studies undertaken by small teams or individuals have evolved into what Derek de Solla Price named *Big Science*, where multidisciplinary teams conduct large-scale studies using expensive resources (de Solla Price, 1963). As a consequence of these two factors, new forms of managerialism have been introduced into the higher education system (Morris, 2002). These have led to an increasing interest from funding bodies in developing external evaluation processes to allocate funds instead of the traditional peer review system (Hicks, 2009).

This section analyzes the changing role of universities and the growing interest on developing quantitative indicators of their performance. It briefly describes different national university models and systems, focusing on their differences regarding the control exerted by public and private parties. Then, it describes the evaluation schemes developed according to these different systems and models and the way they have expanded in different parts of the world. The focus will be on the role played by the field of bibliometrics as a plausible option for the development of 'objective' indicators of research performance. Next, it discusses the consequences of performance-based research funding systems (hereafter PRFS) in the higher education system, and the challenges that research policy makers, researchers and bibliometricians are currently facing. Finally, it will look forward to the next section by referring to the rise of a global higher education landscape which blurs differences between national systems.

## The evolving mission of universities: Theoretical University models versus actual University systems

The development of universities as educational and research institutions is a history of controversies, ideals and the constant revision of concepts and goals due to discrepancies on what their main role is. Initially designed as elite centers for highly qualified education, they have evolved into complex institutions that combine traditional roles inherited by their own history with the demands of current times that impose flexible and dynamic entities capable of generating wealth to benefit their context. The dramatic changes research has experienced during the 19th and 20th centuries have not only influenced the way people live and perceive reality, but they have also had deep consequences on how science is produced. The utility of science and the benefits from research are evident, building on an unprecedented reputation of the scientific enterprise. However, the expenses entailed are higher than ever, requiring financial support for equipment, staff and training.

Such phenomena have produced deep changes in the mission and expectations universities arouse in society. Their historical role as education providers has prevailed for centuries, originally leaving research and scientific progress to other institutions such as national societies and academies (Manjarrés-Henríquez, 2009). This is the case of the United Kingdom, where the Royal Society channeled scholarly communication within the research community. Another variant is that set by the French Academy of Sciences, which adopted a key role coordinating and controlling research development in France and expanding its model to other European countries.

But it is at the end of the 19th Century, with the rise of the German universities (Barnes as cited by Torres-Salinas, 2007) when research enters higher education, establishing itself as the second mission of universities. This role has grown over time and universities have become the main motor of scientific progress. This highlights their potential as generators of wealth and has entailed their shift from the ivory towers in which they were embedded to the heart of society. This new concept of university as a research and educational institution is known as the German or Humboldt's Model. Etzkowitz (1990) refers to this change as the first academic revolution, which will lead to a need to assess and evaluate the research performance of universities. A once minimal interest in student satisfaction turns into a real concern to understand and monitor the behavior of universities in order to optimize and assess the production of knowledge and the human capital involved (Geuna, 1999).

Hence, two different variants of higher education models are identified, expanding throughout Europe and North America. The first one would be the French or Napolean Model, where a national entity keeps its leading role as coordinator of the country's research development (the Spanish system developed in a similar fashion until the 1970s with the establishment of the Spanish National Research Council). The second one is that set by the German universities, where the government takes charge of developing a legal framework while universities are responsible for the development of scientific progress. Here, although the Max Planck Institutes play a leading role in research, they work independently to universities. This model will later lead to different variants depending on the levels of governance universities exert. It will deeply influence countries such as the United Kingdom or the United States, where the importance of universities as generators of scientific knowledge has prevailed. As a result, there is a constant tension between the State and the universities. The former, the main source of universities' funding, tries to direct them as a tool to achieve prosperity. On the other hand, the universities are immersed in an identity crisis trying to maintain their independence but also knowing their importance as key players in the economic and political development of their countries.

The second mission of universities was finally favored over teaching as research advancement became an icon of nationals' supremacy. Two views have been adopted. First, the scientific ethos set by Merton, which defines the theoretical motivations of researchers to pursue scientific knowledge. Second, the idea that scientific progress will lead to technological and industrial knowledge (Bush, 1945). This latter notion has lately expanded with an on-going discussion of a third mission of universities (Nedeva, Barker & Osman, 2014). Here, universities are not only expected to generate knowledge and offer academic training, but also to apply and exploit this knowledge outside of academia, having a direct impact in society (Laredo, 2007). At this point, a third player emerges: the market. This new context shapes a landscape in which countries develop distinctive national university systems determined by the relations between market, state and university.

*Figure 1. Adaptation of the Triple Helix by Etzkowitz & Leydesdorff (2000) to the theoretical university models*

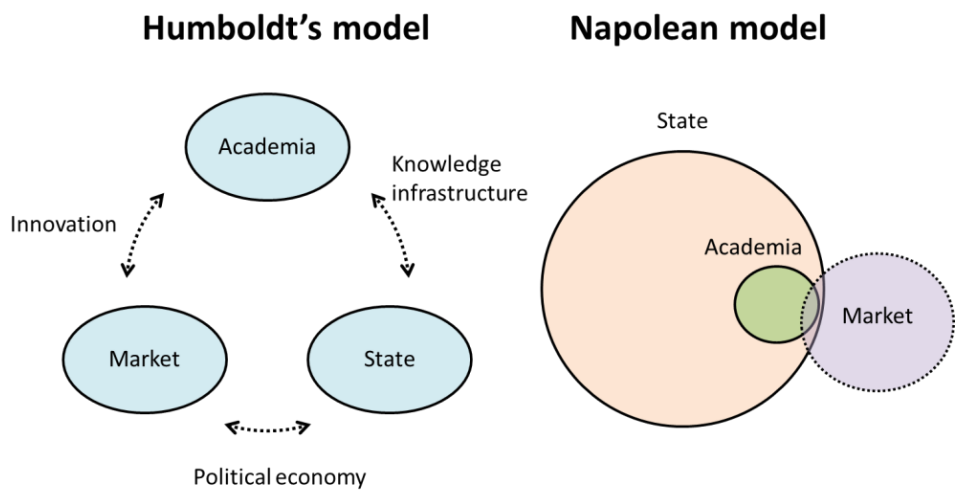

These tensions are presented through the Triple Helix theory formulated by Etzkowitz and Leydesdorff (2000). This conceives of the relation between these three entities as a network in which each actor reinforces and influences the other two, establishing an 'endless transition' by which knowledge is generated and then used as a source for production and distribution. The increasing importance of knowledge in the innovation process introduces universities into a system in which the only actors at the time were market and the state. This introduction involves a series of processes that transform the way production, information exchange and the use of knowledge take place. According to (Etzkowitz, 2000) these processes would be: 1) an internal transformation of each institution with regard to the way in which they interconnect, 2) a greater influence of each institution over the other two actors, 3) the emergence of new entities resulting from the institutionalization of the established networks and 4) a drift from the disinterested generation of science as conceived by Merton (1973) to a science based on goals. As observed in figure 1, the Napolean university model neglects the role of market from the system relegating it to a secondary actor and creates a dependency relationship between academia and state. The triple helix adapts much better to Humboldt's model where universities have greater levels of autonomy, allowing them a much closer interaction with the market and a better position when negotiating and relating to the state.

As a means to adapt and facilitate the interaction between these three actors, since the 1980s most countries have introduced PRFSs into their system. Their aim is to allocate funds based on the research performance of academic institutions. Two countries have set example in this regard: the United States and the United Kingdom. The first one was a pioneer in the development of non-governmental university rankings, highlighting the publication of the 1st edition of the America's Best Colleges University Ranking in 1983. The latter launched in 1986 the first Research Assessment Exercise, an example of governmental action to monitor research performance. At the time, both countries were reaching the end of the huge economic recession that followed the end of the 'Cold War'. With the conservative parties in power, Reagan in the United States and Thatcher in the United Kingdom undertook important reforms in all sectors, including Higher Education. It was a time to reflect when public funding had to be justified by any means and return on investment became a priority. Hence, it is not surprising to see the emergence of a highly competitive environment in which universities had to prove their worth. These two events flag the beginning of two initiatives that began separately but which illustrated the response of government and market to the economic constraints, and the social pressure they would ultimately put on universities.

## The introduction of national multi-university research evaluation systems

While governmental PRFSs saw an earlier expansion, especially during the last decades of the 20th Century, the globalization of Higher Education, the massification of universities and the emergence of an entrepreneurial university model led to the final outbreak of world-class university rankings at the beginning of the 21st Century. For the sake of clarity, here we will follow the chronological order in which each evaluation system expanded.

According to Abramo, Cicero and D'Angelo (2011), national agencies and governments implement such evaluative measures by two possible routes which are not exclusive and may co-exist. The first one has to do with the introduction of national Research & Development programs for evaluating research project grants (i.e., the Spanish National Research & Development Plan described in Cabezas-Clavijo et al., 2013). The second and most widely extended channel takes place through the implementation of PRFSs. Most of these are focused at the institutional level and take place periodically. The first country to introduce these systems was the United Kingdom. It is normally implemented in national university systems where there is internal competition within institutions to attract talent and funding, leading to a stratified system. Here, agencies assess institutional performance and universities are the ones in charge of the evaluation of their personnel. However, there is another variant which focuses on the evaluation of individuals at a national level, adapting the British model to non-competitive higher education systems where researchers may be widely dispersed throughout universities (Abramo et al., 2011; Jiménez-Contreras, Moya-Anegón, Delgado López-Cózar, 2003). Here the focus is made upon PRFSs centered on the institutional level as these evaluation schemes establish a direct link to the current demand for university rankings in the research policy context.

The first system implemented with these characteristics and probably the one which has gained the most attention is that undertaken in the United Kingdom: the Research Assessment Exercise (RAE, 1989-2008), now converted into the Research Excellence Framework (REF, 2014). The RAE has been subjected to many studies describing its methodology and evolution (e.g., Barker, 2007), analyzing its advantages and disadvantages (e.g., Clerides, Pashardes & Polycarpou, 2011; Geuna & Martin, 2003) or discussing its effect on researchers and universities (e.g., Elton, 2000; Moed, 2008; Yokoyama, 2006). Similar experiences have been reported in countries such as Australia, The Netherlands, Norway or Denmark for instance (Auranen & Nieminen, 2010; Geuna & Martin, 2003; Hicks, 2012). The role played by bibliometrics when implementing them has been undisputable (Hicks, 2009), leaving traditional peer review in the background. However, there seem to be mixed feelings about results (Anon, 2010) and contradictory views as to the effect these evaluation systems have had on individual researchers' performance (Abramo et al., 2011). The main problems have to do with the misuse of bibliometric methodologies, a perceived mistrust of peer review among policy managers and a simplistic, deterministic view of the effect economic incentives have on research performance and on the transfer from basic research to innovation. The bibliometric community has often warned of the limitations of bibliometric indicators, indicating the need for good practices, quality data and to combine these methodologies with peer review (i.e., Moed, 2007; van Leeuwen, 2007). However, these have mostly been ignored, turning the use of bibliometrics as a research policy tool into a controversial issue.

The use of bibliometric indicators leaves a sour taste in mouth of the research community, who are not convinced by seeing their performance reduced to numbers (Abbott et al., 2010). In contrast, research managers are relatively satisfied with their use, as they provide 'objective' measures which seem easy to interpret. This allows them to partially remove or at least combine them with peer review, which is seen with apprehension (Schneider, 2009). This replacement can be seen in the evolution of the criteria used to evaluate research performance in many countries (Hicks, 2009). As far as bibliometricians are concerned, they believe in the potential of bibliometrics as a research policy tool but are not at ease with the heavy dependency PRFSs have on metrics, which they consider should be used as a support tool and not as the key criteria (as noted in the first principle of the Leiden Manifesto by Hicks et al., 2015). In general terms, PRFSs seem to be beneficial at first until the costs to maintain them surpass their benefits (Geuna & Martin, 2003). The implementation of these systems is extremely expensive, especially when they include peer review. In the case of the United Kingdom, Martin (2011) considers that the RAE lost purpose and meaning right after the third evaluation took place. PRFSs have an immediate effect on the budgets of universities affecting researchers' resources and scientific careers. This is why they feel very strongly about them, involving themselves in heated debates on the advantages and disadvantages these systems offer. In principle, these exercises should be perceived positively, as they are based on meritocratic criteria. Also, they represent a good opportunity to link research with policy, which makes it easier to argue in favor of the need for research investment when facing funding bodies. But in order to have an efficient system, it must be relatively cheap, transparent and constantly evolving to reflect the changing needs of national university systems. So far, performance-based research systems have benefited the countries that have implemented them, increasing their research output and visibility (Jiménez-Contreras et al., 2003; Martin, 2011; Moed, 2008) but there is a perceived urge to reformulate them in order to keep them useful.

## THE RISE OF INTERNATIONAL UNIVERSITY RANKINGS BASED ON BIBLIOMETRIC DATA

Current bibliometric international rankings are a projection of this new managerialism perspective of research as a countable activity. In principle, it should be the perfect tool for supporting research policy making: it is much cheaper than PRFSs, and it adapts nicely to the emergent global higher education landscape. However, it ignores researchers' concerns on relying solely on bibliometric indicators, and what is more, bibliometricians' warnings on their limitations. Although rankings are quite common in bibliometric studies, university rankings have not considered the use of bibliometric data until recently. Moreover, despite their success in the United States, they have had little presence in the European research policy scenario. Their success can only be understood when looking at the bigger picture and analyzing the evolution of the higher education landscape during the second half of the 20th Century. Here we summarize the main key factors which may help to understand their wholehearted adoption as research policy tools:

- **The globalization of the research-oriented model**. The increasing importance given to the so-called second mission leads to the emergence of a university model focused on research excellence. This enhances the globalization of higher education in which technology, knowledge, people, and ideas flow across national borders.

- **The internationalization of universities.** As a consequence of globalization, universities adopt an international model for promoting joint ventures and a race for talent developing international recruitment strategies.

- **Diversification of funds.** Governmental dependence weakens as universities adopt a diversified funding model, searching for private as well as public investment.

- **Massification of education.** The expansion of the mass education model from primary and secondary education reaches higher education which has to combine top research with massified teaching.

These factors lead to an 'Emerging Global University Model' which seeks to develop World-Class Universities which compete on an international scenario, ignoring all national barriers (Mohrman, Ma & Baker, 2008). These institutions represent the elite, characterized by their strong economic position, resources and reputation very far ahead from the rest of the universities. They have great levels of autonomy and are at the frontier of research and teaching. These elite is formed by up to 100 universities at most; as argued by Billaut and colleagues (2010), the distribution of the global normalized score of the 500 universities included in the Shanghai ranking is highly skewed, with an almost flat curve after the first 100 institutions.

International rankings have become the main yardstick for benchmarking these new research-oriented super-universities in research policy (Baker, 2007). The first ranking launched was the Shanghai Ranking in 2003, followed the next year by the Ranking Web of Universities (Aguillo, Ortega & Fernández, 2008) and the joint version of the THE-QS Ranking, which in 2007 split into two independent rankings. The NTU Ranking appeared that same year, originally developed by the Higher Education Evaluation and Accreditation Council of Taiwan and since 2012, published by the National University of Taiwan. Figure 2 shows a brief chronology of the evolution of the different evaluation schemes.

*Figure 2. Timeline with the main milestones in the development of university rankings and performance-based research funding systems. In yellow, the most significant milestones referring to rankings*

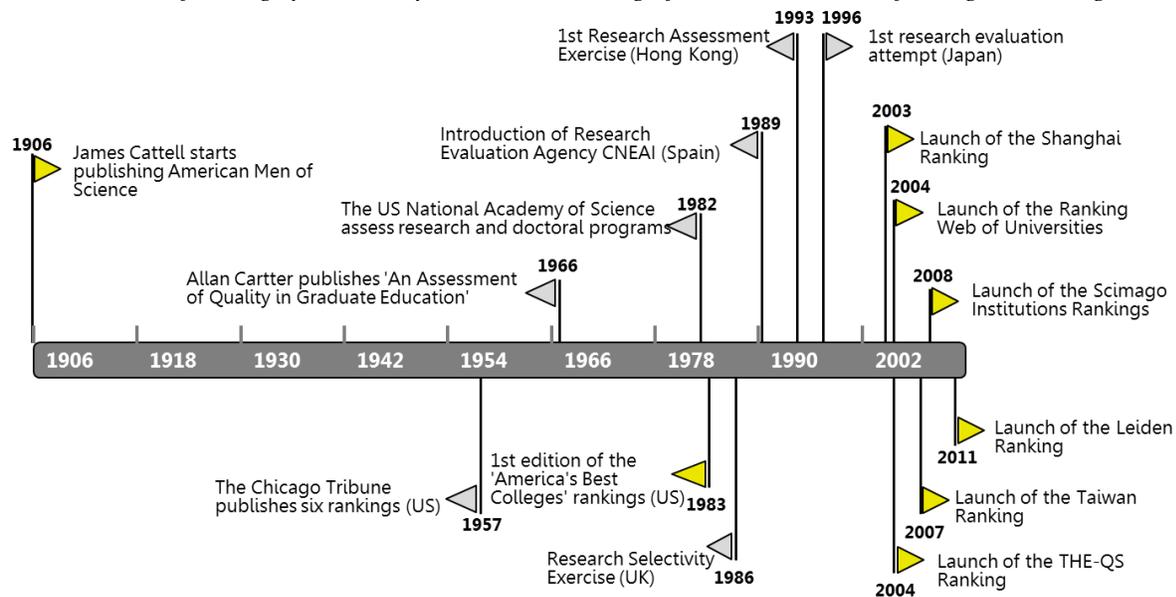

University rankings soon gained a great popularity among research policymakers, stirring a shift on the position held until then by the bibliometric community. Soon they understood that, despite their criticisms and discouragement (i.e., Ioannidis et al., 2007), rankings' popularity continued increasing. Hence, in the subsequent years highly prestigious centers and research groups in the field of bibliometrics launched their own rankings and in 2008 the Leiden Ranking appeared followed by the Scimago Institutions Rankings. However, the former was published at irregular intervals and it was not until 2011 when it adopted its current annual frequency.

Along with the NTU Ranking, these two rankings have certain characteristics that differentiate them from the others. First, they focus only on research, the second mission of universities, acknowledging the inability to rigorously measure other dimensions of universities such as teaching or innovation. Secondly, they offer a very restrictive definition of what is considered research performance. Hence, research output is defined as journal articles mainly indexed in the corresponding database used for data retrieval produced by a given university during a fixed time frame. A good example of such restrictions is the complete name of the NTU Ranking: Performance Rankings of Scientific Papers for World Universities. Finally, regarding the use of a global indicator, the NTU Ranking is the only one of these three which employs it, while the other two include a battery of indicators acknowledging the impossibility of compressing into a single number the research performance of an institution. Once again, this shows discrepancies as to which methodology and choices are better.

Still, there are other technical issues which are acknowledged in many cases by the authors of bibliometric rankings and are difficult to avoid. Waltman and colleagues (2012) mention some of them when presenting the Leiden Ranking. These have to do with the collection of data which is mainly based on the address field of bibliographic records and with methodological decisions made for calculating the indicators. Here we summarize them:

1. **Data retrieval.** Not only institutional name changes and restructuring may affect the quality of the data retrieved, but also the lack of normalization of this field may lead to false positives (publications mistakenly assigned to a given institution) and false negatives (publications not assigned to the right institution). Robinson-García and Calero-Medina (2014) thoroughly analyze the address field in the Web of Science database offering a complete description on the many problems one may encounter when using this field for evaluation purposes.

2. **Methodological decisions.** Other differences between bibliometric rankings have to do with the battery of indicators shown as well as some methodological decisions taken when calculating them. These decisions have to do with the counting of publications (full or fractional counting), the inclusion of document types or coverage differences derived from the use of the Web of Science or Scopus databases.

## Methodological compromises and their reliance on bibliometric databases

In table 1 we include a list of the main rankings indicating the launch year, frequency of publication, type of data and source used to retrieve the bibliometric data and the weight bibliometric indicators have in the final score. As observed, all of these rankings rely to a greater or lesser extent on bibliometric databases and the final score universities receive is heavily influenced by their research performance. This section analyzes two specific aspects related to such reliance on bibliometric data and scientific databases and how ranking producers have developed alternative league tables to partially surpass such limitations. First, a description on the coverage limitations of Web of Science (Thomson Reuters) and Scopus (Elsevier) towards certain scientific fields will be offered. Any research analysis dependent on these databases at the institutional level must be performed with care. Although this is something well-known by the bibliometric community (van Leeuwen, 2007), it is problematic when producing university rankings. It is not possible to verify with the institution in a systematic and relatively effective manner that all research output is considered, or even the share of the total research output which is indexed in these databases. However, it has been extensively reported the limitations these two databases have towards the fields of Social Sciences and Humanities as well as Engineering (Moed, 2005; Moya-Anegón et al., 2007).

Soon criticisms to rankings centered their attention on the disadvantage specialized institutions in fields misrepresented in these databases had in rankings. As a response, university rankings started to incorporate rankings by fields in 2007 (Liu, 2015). Also this allowed ranking producers to offer a more enriched picture of higher education systems and defend themselves against criticisms of the over-simplistic view of universities they offered. Still, this is a partial solution as it may lead to misinterpretations as rankings by fields rely on bibliometric subject classifications and do not reflect the organizational structure of universities, experiencing a mismatch between what it is expected to see and what is actually shown in these league tables (Robinson-Garcia & Calero-Medina, 2014). The second part of this section will focus on such issue.

*Table 1. General description of the main international university rankings*

|  | Launch year | Type of data | Bibliometric data source | Weight of bibliometric indicators |
|---|---|---|---|---|
| Shanghai Ranking | 2003 | Bibliometric and reputational | Web of Science | 90% |
| THE World University Rankings | 2004 | Bibliometric, surveys and manpower | Web of Science | 60% |
| QS Top Universities Rankings | 2004 | Bibliometric, surveys and manpower | Scopus | 40% |
| Ranking Web of Universities | 2004 | Bibliometric and webometric | Scopus | 16.7% |
| NTU Rankings | 2007 | Bibliometric | Web of Science | 100% |
| Scimago Institutions Rankings | 2009 | Bibliometric | Scopus | 100% |
| Leiden Ranking | 2011 | Bibliometric | Web of Science | 100% |

## Web of Science and Scopus: A partial view of the scientific world

The Web of Science database (hereafter WoS) has been historically considered and used as the main data source for bibliometric analyses. The characteristics of its journal citation indexes (Science Citation Index, Social Sciences Citation Index and Arts & Humanities Citation Index) relating cited with citing documents and its renown journal rankings based on the Impact Factor (Journal Citation Reports) have been considered unique until the beginning of the 21st Century. The Science Citation Index, first of its kind, was devised and developed by Eugene Garfield, one of the founding fathers of the field of bibliometrics, in 1963 (Garfield, 1964). Following the example set by the citation index for the field of Law developed by Shepard in the late 1870s, it gave future bibliometricians a basic resource from which to explore the potential of bibliometric data. As Wouters (1999) states in his review of the creation of the Science Citation Index:

*It promised to make an old dream come true: the application of "the scientific method" to science itself, an idea central to the science of science.*

In 2004, a year after the launch of the Shanghai Ranking, the leading scientific publishing firm Elsevier, released Scopus, a scientific database with similar capabilities and characteristics to that of WoS. Scopus has a wider coverage of journal publication output than WoS (Meho & Yang, 2007; Falagas et al., 2008).

However, its journal inclusion criteria are considered to be much less rigorous towards journal quality than the former (Taşkın, et al., 2015). Regarding other types of data used for ranking universities, the Shanghai Ranking includes reputational data based on Nobel Prizes or other field medal awards. The THE World Universities Rankings and the QS World University Rankings both use survey data to determine the quality of teaching or the perception certain so-called experts have of universities. They also use national statistics when available in order to determine the number of international students for instance and other similar variables. Finally, we find that the Ranking Web of Universities uses webometric data. The reason for this lies in the purpose of this ranking which is not to benchmark the perceived quality of universities, but to encourage Open-Access and web presence.

Hence, if we ignore this latter ranking which has a completely different nature, we observe two types of rankings: those which aim to analyze all dimensions of universities and those which only focus on the research mission of universities. We also observe that this conceptual difference follows a chronological order, establishing an analogy with the shift from peer review to bibliometric analysis in research evaluation (Hicks, 2009). In this case, bibliometric rankings represent the response of the bibliometric community to the emergence of the first international university rankings which were heavily criticised by the former. In this context, the selection of WoS and Scopus as their main data sources is expected, as these databases represent the main citation indexes for international scientific literature (Gavel & Iselid, 2008). But still, both databases have important drawbacks which can affect the bibliometric indicators of certain universities and/or countries giving a false image of poor performance. These shortcomings are resumed here:

1. **Language biases towards English-language literature.** Both WoS and Scopus are known for being heavily biased towards research literature written in English language (Moed, 2005; Moya-Anegón et al., 2007). This means that non-English literature is not only misrepresented, but also has a citation disadvantage. Although this should not compromise the results for basic research, but it does represent an important drawback for professional-oriented literature such as Clinical Medicine or fields where national language have a greater presence such as the Social Sciences or the Arts & Humanities (van Leeuwen et al., 2001). It also positions English speaking countries in a better starting point towards non-English speaking countries.

2. **Field biases towards journal publications.** Again, both databases are mainly focused on journal publication data. Journal articles are considered the main communication channel in most fields (Research Information Network, 2009) but the fields of Social Sciences, Humanities and Engineering (Larsen & von Ins, 2010). While monographs play an important role for the two former, the latter publishes an important share of its output through proceedings papers. This means that the output of universities in these fields is highly misrepresented in university rankings.

Recently, Google Scholar is being pointed out as an alternative data source for bibliometric analysis as it surpasses many of the limitations aforementioned (Martin-Martin et al., 2015). However, there are still serious drawbacks that cannot be avoided for institutional evaluation, furthermore for the development of university rankings (Delgado López-Cózar, Robinson-Garcia & Torres-Salinas, 2014; Prins et al., 2014). As a consequence, the use of bibliometric indicators for developing rankings in these areas remains unsolved.

## Bibliometric subject classifications and the organizational structure of universities

The first ranking to incorporate league tables by fields was the Shanghai Ranking which included five broad fields in 2007 and five more rankings in specific fields in 2009. Currently most rankings include tables by fields with the exception of the Scimago Institutions Rankings and the Ranking Web of World

Universities. These league tables represent a major improvement as they provide an enriched view of the performance of universities (Robinson-Garcia et al., 2014). University rankings by fields disaggregate universities' research output based on the premise that comparisons will only make sense when we measure the same type of research output. In principle, such approximation should be sufficient as a means to promote institutional diversity as it would allow specialized universities to excel in their fields of endeavor. Also it should minimize the lack of coverage of databases in certain fields as offering rankings by fields would mean the share of omitted literature would be uniformly lost for all institutions.

But the technical and methodological choices to construct such fields along with a lack of transparency on how these fields are constructed threaten their reliability and lead research policymakers to misinterpretations. Ranking consumers expect to see in these fields a reflection of the research performance of different organizational units of their institutions. This makes sense as it would offer a unique insight to policymakers allowing them to compare certain units with similar ones belonging to other universities (i.e., faculties, departments, etc.), and in fact it is the ultimate goal of such solutions (Bornmann, Mutz & Daniel, 2013). However, bibliometric data is retrieved using a top-down approach. This means that the university output is identified by searching in these databases in the *address* field of each record for the different name variants of a given institutions. Such approach imposes serious limitations in the data retrieval process (Waltman et al., 2012) as not only institutional name changes and restructuring may affect the quality of the data retrieved, but also the lack of normalization of this field may lead to false positives (publications mistakenly assigned to a given institution) and false negatives (publications not assigned to the right institution). Robinson-García and Calero-Medina (2014) thoroughly analyze the address field in the Web of Science database offering a complete description on the many problems one may encounter when using this field for evaluation purposes. They conclude highlighting the impossibility of using address data to develop rankings by organizational units.

In order to surpass this limitation most rankings, or at least those which have reported the methodology followed to construct their classification system, base such system on the aggregation of journals into categories. For example, those rankings based on bibliometric data from WoS will construct their fields and disciplines by aggregating subject categories from the Journal Citation Reports (Robinson-Garcia et al., 2014). This way a proxy to the structure of universities is shown by the premise that researchers specialized in a given field will publish in journals of such field. While this limitation does not question the validity of rankings by fields, it has important consequences towards its interpretation and the potential use that research policymakers can make of them (Mutz & Daniel, 2015). In this regard, two aspects are highlighted: 1) the lack of transparency shown by many rankings towards the construction of such fields, and 2) the different solutions given by those rankings which share the methodology followed for the construction of such fields.

## THE COMPLEXITY OF DISCIPLINARY PROFILES

Despite their efforts to improve their quality both, rankings and evaluation exercises, impose a serious threat to institutional diversity (Lopez-Illescas, Moya-Anegón & Moed, 2011; Lee, Pham & Gu, 2013). Such threat has not too much to do with the technical limitations or shortcomings they may have; but with the simplified view they end up offering and which is the one research policymakers are demanding in many cases. Such construct of the reality, no matter how rigorous it may be, if considered carelessly tends to shape and impose a homogeneous behavior among the assessed units of analysis. Although PRFs implement more sophisticated evaluation schemes and despite rankings now offer a wide range of indicators, such threats are still very much alive as suggests the experience had in the United Kingdom and the debate that followed the reform of the Research Assessment Framework to the recently implemented Research Evaluation Framework (Smith, Ward & House, 2011).

The main issues highlighted here have to do with: 1) the effects evaluation exercises have on researchers modifying their publication patterns (Smith, Ward & House, 2011), 2) the suppression of interdisciplinary research by journal classification systems (Rafols et al., 2012), 3) the advantageous position of mainstream topics (Lee, Pham & Gu, 2013), and 4) the advantageous position of basic research as opposed to applied research (van Eck et al., 2013). While the first issue is implicit in any evaluation process and its solution can only come from a proper use of research evaluation tools, the other three are implicit problems in rankings which are unavoidable if one is to maintain the simplicity of these tools. Indeed, universities are heterogeneous and complex institutions which do not necessarily pursuit the same goals or are defined by the same ground rules (Collini, 2011). Of course this does not only affect to their disciplinary focus, but when focusing on the evaluation of research, it does have a clear impact on universities positioning (Visser, Calero-Medina & Moed, 2007).

In this section we present the main contribution of this chapter, that is, the journal publication profile methodology as a means to visualize institutions and identify clusters of similarity within fields. Although such methodology does not avoid the use of a classification system completely, it adopts a multidimensional perspective as to what a university is in terms of disciplinary diversity. Instead of developing a classification of institutions, it maps them through network analysis showing through the proximity of the nodes the (dis)similarity between institutions in a given area. In order to present this novel approach, we will first discuss the literature regarding previous efforts at identifying institutional profiles. This will offer a brief review to the reader as an introduction to the subsequent description of the methodology presented.

## Methodologies to classify universities by type

Reducing to a single measure the activity of universities is a simplistic view even when focusing on a single dimension such as research. Even if they offer a battery of indicators instead of a single one, these do not allow the reader to see the type of research each university is producing and how they are performing in comparison with other institutions of similar characteristics. The main issue here is how to define which the institutions with similar characteristics are. In order to surpass such limitation, many have suggested the development of institutional classifications. While rankings by fields offer a fragmented view of universities' research output, global rankings neglect institutional diversity; hence it seems reasonable to establish university categories. But any classification must follow a set of criteria by which to discern which characteristics define as (dis)similar two institutions. Once these criteria are defined, the indicators for measuring them should be established. Also, such classification should be straightforward, transparent and universal. If these premises are given then university rankings by institutional type can be offered and institutional diversity will be preserved.

There are already many authors suggesting such approach. For instance, Shin (2009) employs Hierarchical Cluster Analysis to classify South Korean universities. He develops a mission-level classification based on research performance establishing five different types of university. However he warns against the heavy reliance of research managers on his classification as it is sensitive to disciplinary specialization. Ortega, López-Romero & Fernández (2011) perform a similar exercise to classify the 109 research institutes of the Spanish National Research Council. For this, they apply three different techniques (Principal Components Analysis, Agglomerative Hierarchical Cluster Analysis and Linear discriminant analysis). In this case, the resulting classification identifies disciplinary aspects of the institutes and defines three types: technological, humanistic and scientific. What is more, they are able to assign publication practices regarding document types to each institutional profile. This has important consequences as it permits the development of specific research evaluation exercises for each type of institution.

Another perspective is to analyze university profiles and characterize them instead of establishing a classification of universities. This can be done for instance, by developing inverse research profiles by breaking down the subject categories or field areas into the institutions which contribute most to each of these areas according to the overall university system (Calero-Medina & van Leeuwen, 2012). An interesting approach is that followed by Bordons and colleagues (2010), where they suggest using input and contextual data along with bibliometric data in order to analyze and establish institutional profiles as this information will allow us to better capture the characteristics of universities. However, their study is of an exploratory nature and focuses more on explaining the Spanish higher education system rather than on the full development of their promising methodology. All in all, the main problem these classifications have is that they are data-driven. Although extremely useful for research evaluation, they have a difficult fitting with university rankings the classification of institutions as well as the number of classes could differ over time as institutions evolve, preventing temporal stability.

This problem has to do with the nature of the exercise rather than with the methods employed: trying to offer a detailed and full view of universities' performance while at the same time assuring certain levels of assertiveness. This tension between the accuracy of the method and the precision of the result is defined as Duhem's Law of Cognitive Complementary (Rescher, 2006). Duhem states that there is an inverse relation between detail and security; leaving certain levels of vagueness which are indeed, what research policymakers intend to avoid by using university rankings. As a solution to such dilemma López-Illescas, Moya-Anegón and Moed (2011) suggest using graphs and other complementary tools along with rankings in order to provide policy managers with a more precise picture while not having to make any methodological compromises that may bias the results obtained.

## Mapping disciplinary profiles as a complementary tool to rankings

The development of visualization tools for research policy is a research front in itself in the field of bibliometrics (Noyons, 2005). The development of science mapping techniques has been within the interest of the field of research evaluation since its conception. Nevertheless, Derek de Solla Price already envisioned the potential use of publications and citations as their main link to manifest the overall structure of scientific fields (de Solla Price, 1965). Since then and especially thanks to the advances in technology, this specific field has evolved expanding its use, not only focusing on domain analysis (Hjørland & Altbrechtsen, 1995) but also on research assessment (Noyons, Moed & Luwel, 1999). Science mapping techniques allow us to analyze the structure of scientific domains, define relations between the units of analysis or classify and identify research profiles (Ingwersen, Larsen & Noyons, 2001; Shiffrin & Börner, 2004). Visualization techniques offer easy-to-read solutions to rapidly establish the structure of a given set of objects identifying the main elements in it. In this regard, the most used visualization technique is that of social network analysis. Here, the resulting map is defined as a set of elements and the existing relationships between them considering as an element any unit of representation of science such as scientific fields, publications, or researchers (Klavans & Boyack, 2009). They are characterized by visualizing these elements, commonly represented in a two or three-dimensional space, and by matching pairs of elements according to their common characteristics.

Social network analysis strengthens its interpretation allowing us to apply network theory and explain the patterns followed by the model represented in the graph (Vargas-Quesada & Moya-Anegón, 2007). In the case of university rankings, the question remains as to how such relationship between universities is defined. It should consider disciplinary focus, type of research (basic vs. applied) and research intensity, while still being flexible enough as to allow certain levels of ambiguity while being self-explanatory and consistent with the information provided by rankings. Some solutions have already been suggested such as that by Bornmann and colleagues (2014) who suggest the development of a geographic map of research centers of excellent scientific performance on different fields. While their solution allows identifying niches of scientific excellence, it does not characterize them and offer tools to correctly interpret the reasons behind their success. On the other hand, the question on how to identify a unit that links universities in such a broad way without analyzing other external aspects rather than the publication output of universities still remains.

Networks have further benefits, because we walk away from developing indicators and go towards visualization tools, we do not make any statement as to which university profile is best. This is left to the research policy maker who will have to adopt a strategic vision instead of reducing their assessment to good or bad. Combining networks with rankings refines the interpretations made from the latter. The position in a ranking may be explained not because of the performance of the research activity of the university, but because of its nature. However, still the question remains. How can we define a link that is not too constrained to a given type of activity (i.e., coauthorship) and, at the same time is not too ambiguous as to lose focus on the meaning of the network? Here we propose the use of journals as units of analysis broad enough as to encompass other elements rather than an enclosed definition of university class and at the same time rigorous enough as to find a balance between detail and security. The journal publication profile methodology establishes that two universities have a similar disciplinary profile with they publish an importance share of their output in the same journals. This similarity can be defined or explained by a series of complementary reasons:

1. **Specialization.** Scientific journals are considered are optimum unit for identifying fields and disciplines, hence the reliance of bibliometric classifications on journals. Two institutions publishing in a same journal have a common interest on the same research topics.
2. **Collaboration.** Also, two universities may be publishing in the same journal as a result of an intense collaboration between researchers of both institutions, reflecting a direct link between them.
3. **Geographical proximity.** The publication in national journals or non-English language journals means that universities belong to the same research community and, again symbolizes a relation between those institutions.

As shown by García and colleagues (2012) and Robinson-Garcia and colleagues (2013), the journal publication profile of universities may be a useful complement to university rankings as it establishes a balance between the compromise bibliometricians adopt when offering quality tools for research policy use, and the easy-to-read tools demanded by research policymakers. Figure 3 shows an example of a picture of the Spanish higher education landscape in the field of Social Sciences based on the journal publication profile of universities. The size of vertices shows the number of publications of each institution while the color intensity reflects the share of papers published in high impact journals (as defined by their Impact Factor). The width of lines between vertices establishes the intensity of the relation between universities based on the number of papers published in common journals. As observed, the main cluster formed is that of the largest universities and those that are better positioned in university rankings. However, universities worse positioned but with a different disciplinary profile are easily identified. In the example shown we rapidly identify a group of polytechnic universities and a group of universities highly specialized in the field of Economics.

*Figure 3. Example for the output Spanish universities in the Social Sciences based on their journal publication profile*

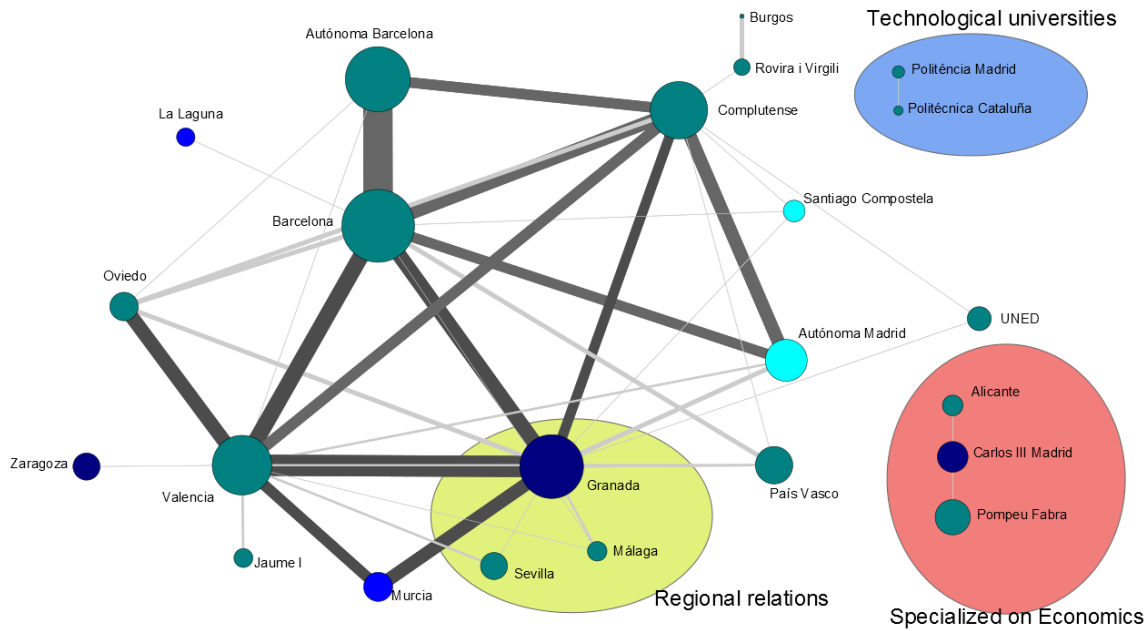

While in this case the color of the nodes is represented by the position of the journals, it could easily be defined by the university's position in a subject ranking. Doing so would allow the reader to explain and understand how specialization affects their position and discern the reasons behind their performance in a given ranking. The potential of such simple tool is very high in our perspective, it informs on the university's profile while at the same time benchmarking it and contextualizing it with other institutions, which is one of the strengths of rankings. It also allows the reader to analyze universities' performance horizontally as well as vertically. While color gives a sense of ranking institutions, position allows the reader to discern with which universities such comparisons make more sense.

## CONCLUSIONS AND SUGGESTIONS

The development of proper methodologies for the elaboration of research rankings is an on-going research front in which many variables and questions still remain unsolved. But still, if we rely on the imperfect picture drawn here, one may question why these tools have had such success with research managers despite their many caveats. The answer is twofold. On the one hand, international university rankings are powerful and persuasive tools for demonstrating university performance to external parties from the university management such as potential students, the media and politicians (Bornmann, 2014). Their interest does not necessarily on their use for developing research strategies rather than for convincing others on the success of previous decisions. On the other hand, international rankings, based completely or partially on bibliometric data, provide accountable and seemingly objective metrics on an international scale. This allows research managers to benchmark their own performance with others, disregarding any methodological concern (Hazelkorn, 2011).

International rankings serve as yardsticks for measuring the pulse of a rising global higher education market. They are the only tools for acknowledging what until their emergence was a blurred world-scale picture in which only certain super-universities stood out (Harvard, Oxford or MIT for instance) while the rest remained hidden (Marginson & van der Wende, 2007). Such visibility accelerates the globalization of a research-oriented university model. They also encourage competition not only at the institutional level, but at the national level, urging governments to offer reliable national statistics and to establish incentives for universities to stimulate competition. But this institutional stratification finally benefits those countries in which universities were already regulated in such a way, that is, mainly the United States and the United Kingdom (Robinson-Garcia et al., 2014).

In this sense, the supremacy of the Shanghai Ranking above the rest as an influential player in the higher education landscape is undisputable. It originally aimed to compare Chinese universities' positions with World-Class universities but, due to the relevance it has gained, it is now used by research managers, students, researchers and the media all over the world. It is also one of the few which has been described in peer-reviewed literature by its authors (Liu & Cheng, 2005). Among others, research managers and national governments have used this ranking to evaluate the health of their national university systems (i.e., Docampo, 2011; Marginson & van der Wende, 2007). This type of exercise aims to rethink and reformulate national university systems, especially from European countries as, in the words of Aghion and colleagues (2010), 'there is little point on promoting competition among universities if they do not have sufficient autonomy to respond with more productive, inventive, or efficient programs'.

The bibliometric community has recently started to discuss the consequences of malpractices by research policymakers of any tool based on publication and citation data, including university rankings. Recently, they came up with a Manifesto that intends to promote good practices among research managers (Hicks et al., 2015). Although the principles promoted in such paper seem reasonable, the evidence a tension between the simplicity of the tools demanded by research policymakers and the desired impartiality bibliometricians wish to offer by showing complex solutions based on a battery of sophisticated indicators.

This chapter discusses the use of a specific technique of science mapping as a midpoint that does not compromises bibliometricians while offering a comprehensible tool that does not substitute but complement university rankings. We consider that the journal publication profile methodology could be a powerful support tool for research policy makers to help them to discern and interpret correctly the information provided by rankings in terms of disciplinary diversity. This is not the first proposal following this line of thought, for instance, the U-Map initiative from the European Union takes a similar perspective. However, because of the many dimensions it intends to capture, it has not been capable of attracting enough attention from university managers. Focusing only on a specific dimension (research in this case) allows us to easily develop mapping tools such as the one described. We believe that drawing the attention on visualization tools rather than on indicators will also force policy makers to adopt a more strategic perspective on to what do they want their university to look, rather than how well positioned they are with respect to others.

**RESPONSE TO REVIEWERS**

February 26, 2016

Dear Dr. Kevin Downing,

Here we include the revised version of the manuscript entitled 'Analyzing the Disciplinary Focus of Universities: Can Rankings Be a One-Size-Fits-All?' in which we take into account your comments and those made by the reviewers. First of all, we would like to thank the reviewers for their constructive comments and criticisms. We agree with all of them and we have made an effort to address them all.

These are the main changes implemented in this new version:

- We have stressed the link between the sections of the chapter and the goals we aim to achieve in this paper. This can be observed in the last two paragraphs of the introduction.
- We have reduced some redundant and marginal parts of section 'From National Research Evaluation systems to International University Rankings'. Especially that regarded with the discussion on the benefits and caveats of PRFSs. We believe they were a bit misleading and, as the reviewers stressed out, it took less importance to the main contribution of the chapter.
- We have revised the style and writing of the manuscript
- We have mentioned the issue of Social Sciences and Humanities explicitly, acknowledging that this is an obstacle that has not yet been solved.
- We have also mentioned U-Map in the conclusions as an example of an initiative that follows the same line of thought we have presented here.